\documentclass[11pt]{article}
\usepackage{titlesec}
\usepackage{titling}
\usepackage{cite}
\usepackage{lipsum} 
\usepackage{graphicx}
\usepackage{amsmath}
\usepackage{amsfonts}
\usepackage{amssymb}
\usepackage{enumitem}
\usepackage{fancyhdr}
\usepackage{authblk}
\usepackage{geometry}
\usepackage{caption}
\usepackage{ragged2e} 

\vspace{-10em} 
\title{\textsc{\fontsize{20}{24}\selectfont\bfseries The Future of Document Verification: Leveraging Blockchain and Self-Sovereign Identity for Enhanced Security and Transparency}}
\author{
  \textsc{\fontsize{13}{14}\selectfont Swapna Krishnakumar Radha,Andrey Kuehlkamp and Jarek Nabrzyski}\\
  \textsc{\fontsize{12}{13}\selectfont Center for Research Computing, University of Notre Dame,Notre Dame\\ Indiana, USA 46556}\\
  \textsc{\fontsize{11}{13}\selectfont\textit{sradha@nd.edu, akuehlka@nd.edu, naber@nd.edu }}

}
\vspace{18pt} 

\date{} 

\begin{document}
\maketitle

\vspace{-5em} 
\section*{\textit{\textbf{\small ABSTRACT}}}
\justify 
\noindent\textit{\fontsize{10}{12}\selectfont
 Attestation of documents like legal papers, professional qualifications, medical records, and commercial documents is crucial in global transactions, ensuring their authenticity, integrity, and trustworthiness. Companies expanding operations internationally need to submit attested financial statements and incorporation documents to foreign governments or business partners to prove their businesses and operations' authenticity, legal validity, and regulatory compliance. Attestation also plays a critical role in education, overseas employment, and authentication of legal documents such as testaments and medical records. The traditional attestation process is plagued by several challenges, including time-consuming procedures, the circulation of counterfeit documents, and concerns over data privacy in the attested records. The COVID-19 pandemic brought into light another challenge: ensuring physical presence for attestation, which caused a significant delay in the attestation process. Traditional methods also lack real-time tracking capabilities for attesting entities and requesters. This paper aims to propose a new strategy using decentralized technologies such as blockchain and self-sovereign identity to overcome the identified hurdles and provide an efficient, secure, and user-friendly attestation ecosystem.}

\vspace{18pt} 

\noindent\textit{\textbf{\small KEYWORDS}}
\\ 
\noindent\textit{\fontsize{10}{18}\selectfont Attestation, Blockchain technology, Self-sovereign Identity technology}

\titleformat{\section}
  {\normalfont\fontsize{14}{18}\bfseries\scshape}{\thesection}{1em}{}
  
\titleformat{\subsection}
  {\normalfont\fontsize{12}{16}\bfseries}{\thesubsection}{1em}{}
  
\titleformat{\subsubsection}
  {\normalfont\fontsize{11}{14}\bfseries}{\thesubsubsection}{1em}{}

\section{Introduction}

Attestation of records is the important first step to verifying and certifying documents to confirm that the information contained in them is correct and authentic and was issued by a legitimate authority. Legal, education and commercial documents often need to be attested when they need to be used for international purposes such as employment, business expansion to foreign countries, and global transactions \cite{attestation}. For instance, in the context of legal documentation, records such as birth certificates, marriage certificates, and power of attorney require attestation to verify and assert the legitimacy of the record when they are used in legal proceedings or administrative reasons in a foreign country \cite{ayub2021educational}. Attested documents prove they can be accepted in jurisdictions outside of the country of origin. In the educational sector, diplomas, degrees, transcripts, and certificates need to be attested if the students wish to study or apply for employment opportunities in a foreign country. An attested educational record confirms that the student's academic achievements and qualifications are trustworthy and have been awarded by an accredited educational institution. Attestation is, therefore, a vital step in confirming the credibility and global acceptance of different types of documents \cite{attestation}.

\noindent
The current manual verification faces numerous challenges and problems. Some of the significant challenges are delays in the completion of the process due to the number of steps, personnel, and time required to complete the process. The traditional system used for the process is also prone to human error, which is difficult to track due to the inability to perform real-time transparent and tamper evident tracking of each attestation step. In addition, the COVID-19 pandemic added difficulty in ensuring the physical presence of the record owners or attesting officers. Another problem is reliance on the traditional system alone to attest paper-based documents, which makes it susceptible to fraud and forgery. There have been several instances of fake educational certificates used for employment or visa applications \cite{ayub2021educational}. In addition, it is often a complex and bureaucratic process, prone to delays and inefficiencies. Geographical limitations can also be a hurdle when individuals in remote areas require attestation from authorities in a faraway location. 

\noindent
This work is organized as follows: The related work section delves into some of the existing systems used for attesting different types of records and how they aim to prevent forgery and reduce time delays involved in the process. The section also explores the system's underlying issues, making the process complex not only for record owners but also for attesting officers and entities across international borders. The proposed system explains how an integration of blockchain and self-sovereign identity technology can be used to create a more interconnected and trustworthy attestation system that can create transparent tracking of the process while preserving the privacy of record owners and the confidentiality of information in the record. 

\section{Related Work}

The attestation process provides a standardized method to verify and certify documents, facilitating global mobility, employment, business, and education. The absence of a formalized attestation process can increase the risk of fake documents being used for cross-border transactions and interactions. Traditional attestation systems involve a multi-step process that is used to verify and assert the authenticity of different records, such as legal documents, educational certificates, and business documents. The primary purpose of attestation is to prevent fraud and forgery, thereby ensuring that the document and the information contained in it are verified by a trusted entity to be credible. 

\noindent
Even though a standardized system exists to verify the documents, fraudulent practices are prevalent, especially in educational qualification documents. This is because, in the era of digital transformations, individuals have several ways to learn and interact with educational institutions. It is possible to complete an entire educational degree through online education, where students can use various online tools to learn and earn their degrees. While this makes education easily accessible to students even in remote areas, it makes it difficult for verifiers to access the credibility of qualification records provided by the students. The increased risk of circulation of fake educational qualification documents can affect the value of many of these educational certificates for courses completed online \cite{alam2021does}.  Recently, Forbes magazine conducted a study on the prevalence of fake diplomas and transcripts, revealing that the degree mill industry is illicitly generating an estimated 7 billion dollars in revenue \cite{ForbesStudy}. An important group of victims is those students who saw the advertisements of these low-quality, for-profit educational institutions and spent their time and money to earn the degree \cite{ForbesStudy}. Fraudulent practices are not the only challenge in the traditional attestation process. The complex document requirements can also be overwhelming for requesters who need to get their foreign public documents legalized \cite{hartoyo2019hague}. Another significant challenge is the lack of a transparent system to receive communication and regular updates from attesting entities regarding the progress of attestation. This lack of a transparent tracking system becomes a significant hurdle for remote applicants who have no option other than to wait to receive the final update, which is either a successfully legalized document or a reason for rejection. 

\noindent
In order to make this traditional legalization process less complex and quicker, the Hague Conference formed The Convention of 1961, which introduced the use of 'Apostille' to legalize foreign public documents for member countries \cite{hartoyo2019hague}. However, there are still countries that are not part of the Hague Convention, which affects its use in a global context. Countries that are not part of this convention are still required to follow the traditional legalization process.  

\noindent
Many blockchain-based approaches have been proposed for document verification, offering an alternate mechanism to authenticate and manage important records. A blockchain-based enrollment system was proposed by Fernando et al. for the University in Indonesia to solve some of the challenges related to university documents such as certificates and transcripts \cite{fernando2020blockchain}. Authors have proposed a blockchain technology-based enrollment process to monitor and control the enrollment activity of each student so that it is possible to ensure that only valid students who completed the enrollment process can register for courses and get course credits for the registered and completed courses \cite{fernando2020blockchain}. Badlani et al. have proposed an Ethereum and IPFS-based approach to store, retrieve, and authenticate educational documents safely. They have also included tools to efficiently organize the complete examination process and generation of associated results \cite{badlani2022educrypto} using mechanisms such as smart contract logic. Smart contracts have been utilized to safely execute the logic associated with the proposed system and generate a tamper-evident chain of transaction records.

\noindent
Blockchain-based approaches have been proposed for academic purposes and the secure sharing of digital assets such as wills on the blockchain. Crypto-Wills is a blockchain-based system proposed to verify and securely transfer a deceased person's asset to the assigned entity using ERC-20 and ERC-721 crypto tokens. A consensus-based mechanism executes the contract specified in the will, verifies and transfers assets to the beneficiaries \cite{Shah2019CryptoWillsTD}. Gunit Malik et al. have provided a detailed description of how a blockchain-based solution can be used to verify the authenticity of documents that are issued by the Indian Government. They have leveraged the availability of private channels in hyper ledger fabric to achieve the privacy of document data.

\noindent
Forgery and fraudulent practices are also present in high-profile areas such as international trade and business transactions, the legal sector, and financial and banking activities. The complexity and the time-consuming nature of the traditional attestation and document verification process are also widespread in all the aforementioned sectors. So, there is a need for a distributed solution that can transparently and continuously monitor and track document verification and attestation from its starting point when a user requests attestation until the final step, which produces the fully attested document. It is also important that the auditing of attestation steps does not violate the privacy of individuals and entities involved in the attestation process, including attestation bodies and document owners. Confidentiality of document data and user information should also be ensured. Document owners should be able to get status updates via a peer-to-peer secure channel on the progress of the attestation process for their documents. Attestation bodies need to be able to contact the entity or individual who completed the previous attestation step in case additional information is required to perform verification. Once the attestation process is complete, a secure communication channel is desired so that the verifier can contact the document owner or the attestation bodies if they require any additional details. All communications need to be securely and transparently recorded in a verifiable manner.

\section{Proposed System}

This section is organized into four subsections: in the first subsections provides a brief introduction to the unique features of blockchain technology, second subsection introduces smart contracts, third subsection introduces self-sovereign identity technology and micro-credentials. A detailed description on the working of the proposed system and its key features are explained in the fourth section.

\subsection{Overview of Key Technologies and Concepts used in the Proposed System}
\subsubsection{Blockchain Technology}

Blockchain has proven useful beyond the financial sector and has demonstrated its applicability in supply chain, real estate, insurance, voting and governance, and the Internet of Things (IoT). Some of the important features that made blockchain technology useful in sectors beyond finance are transparency, immutability, autonomy, security and also its decentralized nature. Blockchain acts as a distributed ledger in which each transaction is recorded in a block and linked to the previous block, thereby forming a chain visible to all the participants in the blockchain network \cite{bhutta2021survey}. This visibility ensures that every transaction on-chain is open to all network participants for verification, creating a transparent environment for all interactions between blockchain network participants. Anyone on the network can see the details and history of transactions on the blockchain depending on several key factors, such as accessibility, consensus, and control mechanisms. This characteristic of blockchain enhances the accountability and integrity of the information recorded on a chain, making any attempt at unauthorized altering of data easily detectable \cite{bhutta2021survey}. Moreover, the entire blockchain ledger is distributed across multiple computers, referred to as nodes, operating in a peer-to-peer manner. Modifying data on the blockchain would require simultaneous modification of the ledger on more than half of the nodes operating on the network, which is extremely difficult to achieve on large and well-distributed networks \cite{bhutta2021survey}. Consensus mechanisms such as Proof of Work or Proof of Stake are used to perform validation and add transactions to the ledger \cite{lashkari2021comprehensive}. These mechanisms help ensure that all nodes on the blockchain network agree on the current state of the ledger, rejecting unauthorized attempts at modifying data on the network \cite{lashkari2021comprehensive}. 

\noindent
Every transaction performed on a blockchain is cryptographically encrypted and signed. Digital signatures used to sign each transaction help verify its authenticity. The signature can be generated only by the holder of the private key, thereby confirming the parties' identity in the transaction. The use of asymmetric cryptography ensures that the public key part of a cryptographic key pair is visible to anyone on the network, and the private key part is restricted to the owner of the key pair.

\noindent
Data traceability is another important feature of blockchain, making it valuable in financial transactions, supply chain management, and clinical trials \cite{bhutta2021survey}. Every transaction entered into the blockchain network is timestamped. This helps to prevent fraud and counterfeiting because it allows stakeholders to verify the authenticity of transactions at any time to detect any fraudulent activity or to ensure that a product on the supply chain is not counterfeit. 

\subsubsection{Smart Contracts}

Smart contracts are self-executing contracts that contain the terms of agreement between two involved parties represented by lines of code written in the chosen programming language. They can store information, process input data and generate output based on pre-defined conditions represented by functions in the smart contract. Smart contracts automatically execute the contractual agreement when the pre-defined conditions are met \cite{khan2021blockchain}. When executed, they enforce an agreement between two unknown and potentially untrustworthy parties without the involvement of any external party. Therefore, smart contracts provide the blockchain with an automated and secure way to translate contracts on paper to their digital counterparts. To ensure that no unauthorized modification of smart contracts occurs, they are stored on the blockchain network itself. Automatic execution of pre-defined conditions on the smart contract reduces human error and the chance of disputes. Ethereum is the first blockchain platform that started the development of smart contracts \cite{buterin2014next}. Some other blockchain platforms that support smart contracts are NXT, Hyperledger Fabric, Cardano, and Polkadot.

\subsubsection{Self-sovereign Identity Technology}

Existing identity systems are still not mature enough to handle the wide variety of digital identities that are being used for various online transactions \cite{NSAArticle}. They fail to provide users with one of the most desired features: sovereignty over their personal information. Trying to achieve it, many user-centric designs proposed turning centralized identities into inter-operable federated identities with centralized control, allowing users to have some level of control over how and with whom their identity information is shared \cite{principlesofSSI}. Achieving user autonomy is the next important step for identity systems to move closer to \emph{user-controlled identity}.

\noindent
Recently proposed self-sovereign identity (SSI) systems promise to deliver this exact capability: instead of users being at the center of identity systems, self-sovereign identity systems advocate that users should be the rulers of their identity \cite{principlesofSSI}. Self-sovereign identity systems are founded on the principle that users should be central to managing their own identities. However, the goal of SSI systems extends further, aiming to ensure the interoperability of user identities and guarantee user consent whenever their identity information is shared with other parties, such as service providers or other users. SSI-based systems should also allow users to make claims, which can include personally identifying information (PII) or other facts about their capability. These include license information (proving their right to operate vehicles), work authorizations and security clearance. Figure \ref{fig:VCImage} shows the roles involved in a typical self-sovereign identity ecosystem and their interactions. 

\noindent
The main building blocks of an SSI system consist of 1) an SSI-compatible digital wallet and 2) Decentralized identifiers (DID). In addition to securely storing and maintaining digital identities, the wallet held by the identity owner must also be able to maintain a history of all identities issued to their holder. They can also function as agents that facilitate secure peer-to-peer communication with other entities in the self-sovereign identity ecosystem. Another important functionality of the wallets is that they help users to decide how, when, and with whom their identity information is shared and control how much information is shared with requesting entities. Decentralized identifiers are globally unique, verifiable, and persistent identifiers that help users cryptographically prove that they are indeed the ones controlling the identity stored in their digital wallet. 
\vspace{1em} 

\begin{figure}[htbp]
\centering
  \includegraphics[width=0.7\columnwidth,trim={0 5 0 5}]{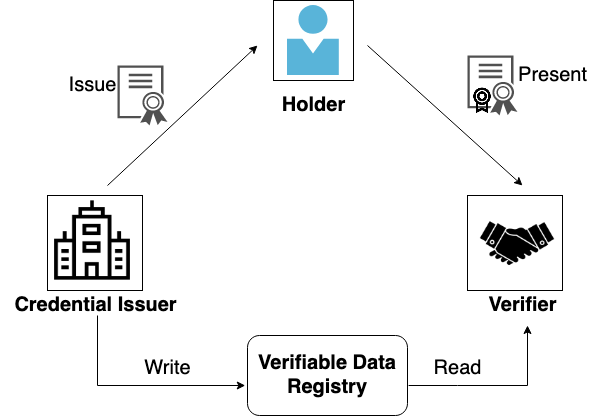}
  \caption{Triangle of Trust for Verifiable Credentials}
  \label{fig:VCTriangleofTrust}
\end{figure}

\noindent
Decentralized Identifiers facilitate key functions for controllers, requesting parties, and subjects. Controllers may be individuals, organizations, or agents managing DIDs. These DIDs enable reliable cryptographic verification of the source of information without third-party involvement, enhancing privacy and autonomy in identity management.
Verifiable credentials (VCs) package all the details related to the DID subject's identity and the cryptographic proof associated with the details into a single file, which can be stored in an SSI-based digital wallet \cite{W3C-VC}. VCs can digitally represent traditional certifications or credentials such as a university degree, driver's license, or certificate of employment. The use of digital signatures makes verifiable credentials tamper-evident. 

\noindent
Figure \ref{fig:VCTriangleofTrust} demonstrates working of verifiable credentials. The first step involves issuer generating a decentralized identifier (DID) containing its public key. The DID along with any other cryptographic material required for the issuer's verifiable credentials is written into a trusted verifiable data registry \cite{triangleoftrust}. Whenever a verifiable credential needs to be issued to a qualified holder, the issuer digital signs the verifiable credential using their private key and sends to the holder who can store it in their own digital wallet \cite{triangleoftrust}. If a verifier needs digital proof of one or more verifiable credential from the holder, holder can generate and return proofs to the verifier. These proof contain the DID of the issuer. Verifier can use it to retrieve the issuer's public key and authorized cryptographic information from the verifiable data registry. The retrieved public key is used to verify and assert that provided proofs are valid and no one has tampered with the proof or the credential \cite{triangleoftrust}.

\subsubsection{Micro-credentials}

According to Pickard, Shah, and De Simone, micro-credentials (also known as micro-degrees, digital badges, or nano degrees) are defined educationally as credentials that encompass the completion of multiple courses by a student yet do not equate to a full educational degree. The educational platform \emph{edX} was the first to introduce micro-credentials in the year of 2013. Following up, many mass open online course (MOOC) platforms now offer different types of micro-credentials \cite{pickard2018mapping}. Micro-credentials allow students to incrementally accumulate certification of skills that they have gained through learning platforms. These types of credentials provide better support for self-regulated learning abilities, allowing students to create learning goals and paths for themselves while still receiving the certification for completed stages \cite{pickard2018mapping}.

\noindent
When powered by SSI technology, micro-credentials can be useful to provide more security and control to these individual pieces of certification.
The benefit of SSI technology lies primarily in its two key components: decentralized identifiers and verifiable credentials. The World Wide Web Consortium (W3C) is currently standardizing decentralized identifiers \cite{W3C-DID}, which will offer a unified framework for their creation and management. Similarly, verifiable credentials, the other fundamental component of this technology, have also been standardized by W3C. These standards provide a framework to generate, issue, hold, and verify these digital credentials in a secure, private, and interoperable manner \cite{W3C-VC}. 

\subsection{Design of the Proposed Approach}

\vspace{1em} 

\begin{figure*}[htbp]
  \centering
  \includegraphics[width=\textwidth,trim={0 10 0 10}]{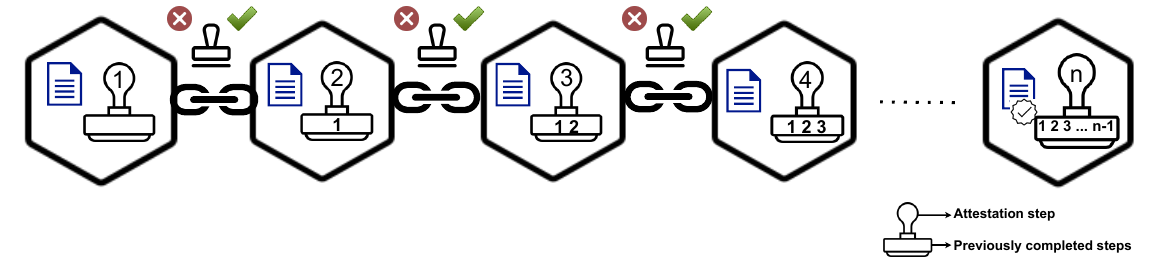}
  \caption{Attestation Chain for a document submitted for authentication}
  \vspace{-5mm}
  \label{fig:chainImage}
\end{figure*}

\vspace{2em} 
\noindent
The proposed system leverages blockchain technology's transparent and tamper-evident nature to generate an attestation chain for each document submitted for verification and authentication. The attestation chain is a secure, tamper-evident, transparent sequence that represents the steps, verifications, and approvals associated with every document users submit for attestation. Figure \ref{fig:chainImage} shows an example of the attestation chain for a document. Each block in the chain shows a distinct phase in the attestation process, capturing non-confidential details related to that step. The primary purpose of the attestation chain is to provide a reliable ledger of the attestation journey that each document has gone through to prove the integrity, authenticity, and legitimacy of the document and process. Blockchain technology ensures that the attestation chain cannot be controlled by any single entity and requires all involved parties to come to a consensus to attest a document. All transactions within each attestation chain are stored on the blockchain and visible to all authorized parties for a clear and transparent view of the attestation process. Each transaction on the blockchain is encrypted and linked to the previous transaction, ensuring a secure chain of custody for the attestation steps. 

\noindent
The system uses SSI technology to generate micro-credentials to represent the successful completion of each attestation step. After completing the final step in the attestation process, a verifiable credential is generated for the document. It links all the micro-credentials together and demonstrates the completion of the attestation process. The following are the major features of the proposed system:

\begin{itemize}
  \item Mobile-based agent for peer-to-peer communication between document owner, verifying entity, and attesting entity;
  \item On-chain transparent tracking and auditing of document attestation steps;
  \item User-friendly interface to view non-confidential details of the document attestation status;
  \item Ability to revoke or expire the attestation stamp on the document;
  \item Ability to track the attested documents that have expired.
\end{itemize}

\subsection{System Architecture}

\begin{figure*}[htbp]
  \centering
  \includegraphics[width=\textwidth]{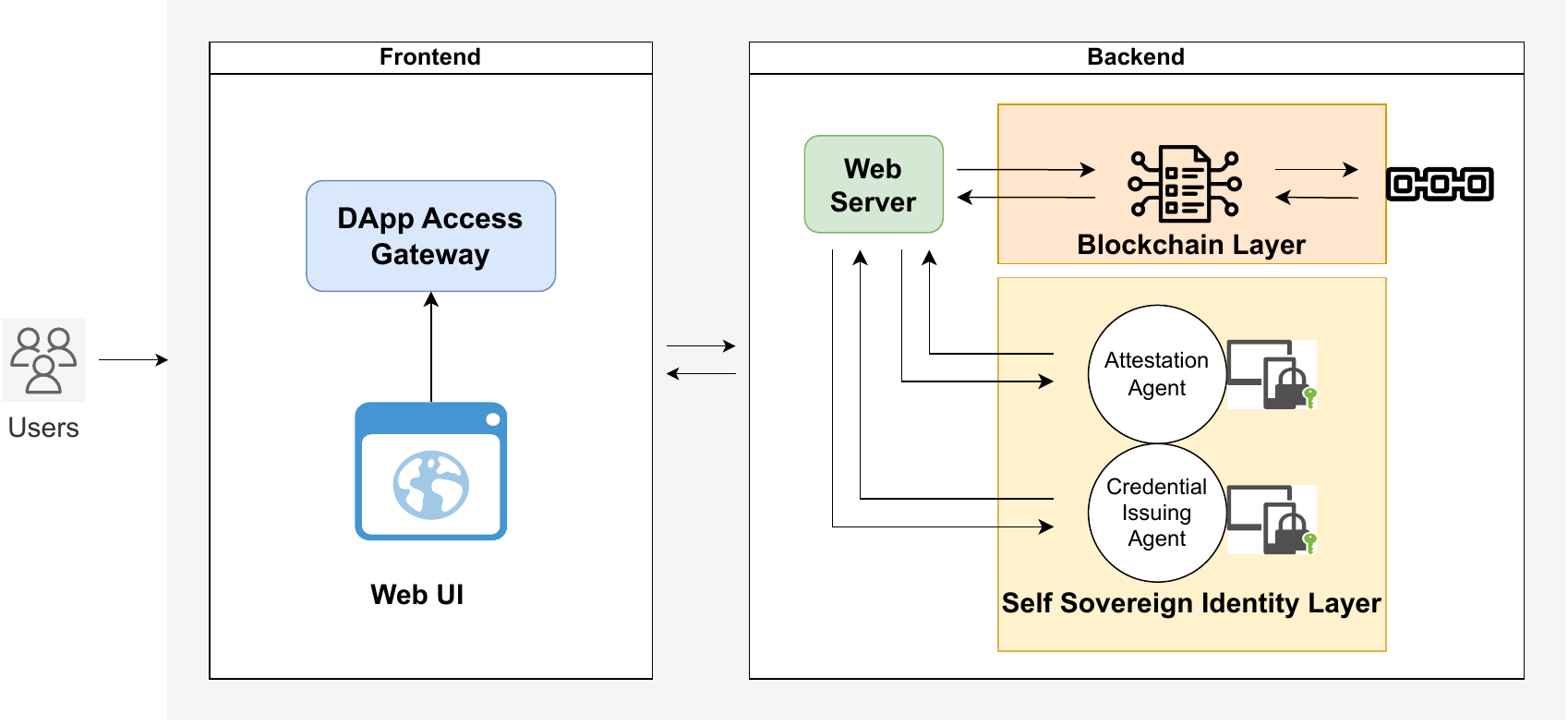}
  \vspace{-8mm}
  \caption{Proposed Architecture Diagram}
  \vspace{-5mm}
  \label{fig:keystepsImage}
\end{figure*}

\vspace{1em} 
\noindent
The system proposed in this paper is designed to function in parallel with the existing traditional attestation process to bring more transparency to the manual process that is being followed currently and enhance the traceability of the entire attestation process. Fig \ref{fig:keystepsImage} shows the conceptual architecture diagram of the proposed approach. A decentralized application (DApp) is used to create and submit blockchain assets, such as attestation requests and non-sensitive details related to each attestation step, which are linked together to form an attestation chain. The DApp Access Gateway is a checkpoint to ensure secure and controlled access to the DApp and the application services. A web server interacts with the blockchain layer to record attestation transactions. The web server interacts with the self-sovereign identity to initiate the creation and issuance of credentials. 

\noindent
Attestation agents and credential issuing agents are self-sovereign identity-based software agents. The attestation agent can communicate with the SSI-compatible wallet, which holds the verifiable credential issued to the attesting entity. The wallet helps to manage cryptographic keys used to sign the micro-credential securely. The keys can also be used to encrypt messages communicated between the attesting entity, users, and verifiers. A credential-issuing agent will be used to generate and issue verifiable credentials to authorized attesting entities, users, and authorized verifiers. The credential issuer will also hold an SSI-compatible wallet holding the verifiable credentials of the issuer. Both the credential issuing agent and the attestation agent will have a DID associated with them, which can be used by users or other verifiers to know the legitimacy of the entity and ensure that they still have ownership of their identity. 

\noindent
The proposed approach involves both blockchain and SSI layers. Therefore, it is important to identify and assign a host layer for each role. Not all roles require identities to perform authorized actions in both layers. 

\noindent
The three key roles in the proposed system are \emph{document holder}, \emph{attesting entity} and \emph{verifier}. The \emph{attesting entity} is a trusted party certified by the government and has the capability to verify and authenticate the submitted document. After verification is complete, the attesting entity is responsible for issuing a micro-credential proving the completion of the attestation step. The \emph{document holder} owns the document, accepts desired credentials from the attesting entity, and holds the verifiable credential associated with the attestation in their SSI-compatible wallet. \emph{Verifier} can be an attesting entity trying to request any additional information from the document holder or an attesting entity from a previously completed step. Employers or government authorities who request proof of attestation or additional information also act as \emph{verifiers}. 

\noindent
The document holder must have unique identities only in the SSI layer after the document has been successfully attested. This identity is issued by an issuing organization, which can be a government entity. This could be a department of government that handles digital identity. For instance, an identity issuer can be an agency that runs a national identity program. The document holder can check the reliability and integrity of the issuing organization using the issuer's DID. The DID will have details such as their unique identifier, public keys associated with DID, privacy considerations followed by the issuer, and details of the revocation registry, if any. A revocation registry is usually used to maintain a decentralized registry of expired and revoked credentials that were issued by the issuer in the past. The SSI-compatible wallet holding the verifiable credential related to the identity of a user will be used to receive notifications regarding the status of their attestation request. The wallet will also be used to hold the micro-credential associated with each attestation step and the final verifiable credential issued to prove the successful completion of the attestation. 

\noindent
The document's unique identifier can be used by the document holder on a user-friendly web interface to check the details related to the timeline of the attestation process that is recorded on the blockchain. The attesting entity requires identification in both the blockchain and SSI layers. Identity on the blockchain layer will be used by the attesting entity to submit a transaction on the chain whenever the required verification is completed by. This transaction will be recorded as part of the attestation chain created for the document. Identity on the SSI layer will be used by the attesting entity to create and issue micro-credentials related to the completed attestation step. Every attesting entity is issued a verifiable credential containing details such as their title, issuing organization's details, issuance and expiration date, cryptographic proof that can be used to verify the credential's authenticity, and public key associated with the attesting entity. The attesting entity can store and maintain the issued credential in an SSI-compatible wallet.  The SSI agent associated with the wallet facilitates peer-to-peer secure communication with other entities to exchange proofs or other necessary information that the wallet owner authorizes to share.

\noindent
Verifiers need not have an identity on the blockchain layer since they can use the document identifier to view the attestation chain related to the document. They require an identity only in the SSI layer. The identity can be used for needed peer-to-peer interaction with the document holder or attesting entity to collect additional information after viewing the attested document. 

\begin{figure}[htbp]
  \centering
  \includegraphics[width=0.7\columnwidth,trim={0 5 0 5}]{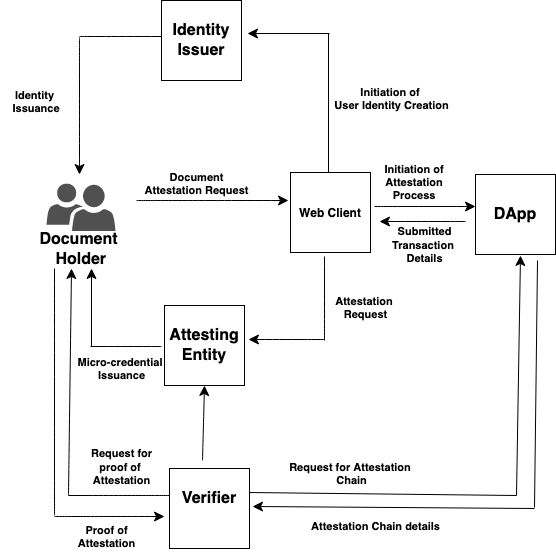}
  \caption{Flow of Information between Entities in Attestation Process}
  \label{fig:flowImage}
\end{figure}

\noindent
Figure \ref{fig:flowImage} demonstrates the steps followed by the proposed system in recording transactions on the chain and generating credentials for user identity and document attestation. 

\noindent
The process starts with a user submitting a document for attestation. The unique document identifier provided as part of the request is used to search and ensure that no duplicate request has been filed for the same document in the past. If there are no duplicates, then the attestation process is initiated by submitting a transaction to the blockchain. The submitted transaction contains details such as the current timestamp, document identifier, and destination country name. The destination country's name is added to ensure that the user's request is not rejected because the document was attested previously for a different destination country where the user pursued higher education or worked. Smart contract logic can be used to ensure that the condition stated above is checked for every incoming request.

\noindent
After the attesting entity performs manual attestation verification and the document is authenticated, the attesting entity uses the SSI agent to generate a micro-credential proving the successful completion of the step. The user can accept and store the micro-credential in their SSI-compatible wallet.  

\noindent
The non-sensitive details related to the completed attestation step, such as the micro-credentials unique identifier, DID of the user, unique identifier of the document, timestamp details, attestation phase number, and DID of the attesting entity involved in that phase, are stored on the blockchain. Smart contract logic can be used to ensure that no step in the attestation process is skipped. After the on-chain verification, the transaction is successfully submitted to the blockchain. 

\noindent
After all the individual steps in the attestation process are complete, a verifiable credential proving the completion of the entire attestation is generated. The entity performing the final step of attestation has the capability to initiate the credential generation process. This verifiable credential will link all the individual micro-credentials, the DID of the entity that initiated the verifiable credential generation, cryptographic proof associated with the issuer, the activation date, and the user's public key. This verifiable credential can be stored in the user's SSI-compatible wallet and can be used to provide proof to the verifier as needed.
\noindent
\vspace{1em} 

\begin{figure*}[htbp]
  \centering
  \includegraphics[width=\textwidth,trim={0 0 0 0}]{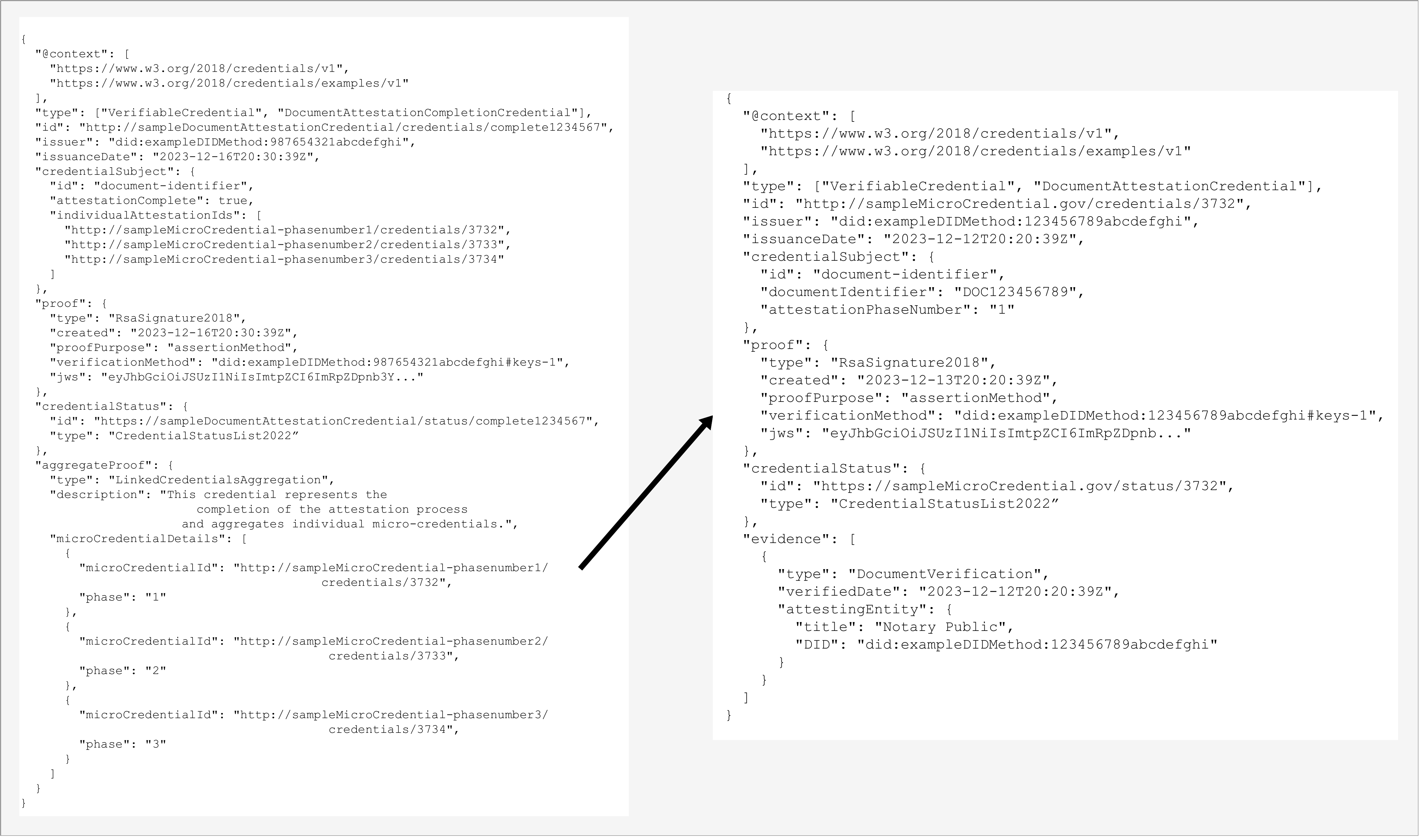}
  \vspace{-8mm}
  \caption{Example Verifiable Credential containing micro-credentials of all the steps in the attestation chain.}
  \vspace{-5mm}
  \label{fig:VCImage}
\end{figure*}

\vspace{1em} 
\noindent
Figure \ref{fig:VCImage} shows an example of a verifiable credential issued to the user, marking the completion of the attestation process. The left-hand side of Figure \ref{fig:VCImage} displays an aggregate verifiable credential, proving that all steps of the attestation have been completed successfully. The CredentialSubject section of the aggregate credential lists the individual identifiers of the micro-credentials, each representing a completed attestation step. The aggregate credential also provides the current status of the credential, which verifies whether the attestation for the document is valid or invalid. It includes the issuer's DID, which can be used to communicate with the issuer securely if additional verification is required. The micro-credential shown on the right-hand side of the figure details one of the micro-credentials issued upon completing an individual attestation step. It provides details about the issuer of the micro-credential, methods for securely communicating with the issuer for additional verification if needed, and cryptographic proof verifying the authenticity and integrity of the credential.

\section{Privacy and Ethical Considerations in the Implementation of Proposed Approach}

A discussion on a new technical solution for handling human data, including personal identification, educational qualifications, work experience, and sensitive health information, is incomplete without addressing its ethical and privacy implications. It is crucial to ensure that the integration of blockchain and SSI technologies does not make any assumptions regarding the ethical and social outcomes of the solution. Advancements in identity management systems must not overlook the potential privacy and security concerns associated with enhancing systems that handle public data. Considering the proposed system as a whole, the rest of this section addresses parameters commonly identified in cybersecurity and data ethics:

\subsection{Accountability}

Accountability can be defined in terms of both responsibility and liability. Accountability requires that there is an identified party who accounts for the way the system works and is capable of explaining the actions and decisions of the system. It encourages responsible behavior from both individuals and entities involved in the ecosystem. It also helps identify who can fix issues when things go wrong \cite{srivastava2021analysis}. Linking transactions in a document's attestation to form a blockchain-recorded attestation chain simplifies audit history access and identifies parties responsible for unauthorized actions during attestation. Using a self-sovereign identity ensures that all interactions between entities are audited securely in their wallet, generating a clear chain of consent and responsibility. Onboarding of all roles into the new system should include educating them regarding their roles and responsibilities in the system.

\subsection{Fairness}

In the context of a blockchain and SSI-based document verification system, fairness can be defined as ensuring that the technology and protocols involved in the system do not discriminate against users within the system. This involves creating and maintaining an environment that provides equal opportunity for all users within the system to access and benefit from it without any bias or discrimination \cite{naik2022evaluation}. Importantly, such a system should not discriminate against users based on their ability to access the technologies. 

\noindent
The system proposed in this paper provides wider access to create, submit, and track information related to the attestation process, regardless of the geographical location. Although internet connectivity can pose challenges in areas with low to no connectivity, the system accommodates this by not hindering the manual document verification process. An offline database can be integrated into the system to record non-sensitive details that need to be stored on the blockchain. When connectivity is available, users can submit their transactions, thereby triggering micro-credential issuance. This ensures continuous operation of the attestation process.  

\subsection{Privacy}

Privacy is the control individuals have over their personal information. In the proposed system, users decide on the data in their micro-credential from the attesting entity, approving or rejecting it based on privacy concerns. This can be done based on the knowledge they gained during the onboarding process about what data is required by the system and how their data is collected, used, stored, and protected. Blockchain is designed to ensure that the system's participants have access to data and metadata stored in the blocks. So, it is important to ensure that data collected and stored on chain does not violate the privacy of any participant in the system \cite{tene2012track}. Attesting entities cannot decide what information is stored on the blockchain. Pre-defined smart contract logic ensures that only necessary and sufficient data is collected and stored on the chain. Periodical auditing of information stored on the blockchain needs to be conducted by authorized entities to ensure that the system complies with the required privacy regulations. A Data Protection Impact Assessment (DPIA) under Article 35 of GDPR (DPIA) is crucial in our system, despite recording only non-sensitive verification data on-chain and storing credential data in user-controlled wallets \cite{DPIA}. Ensuring GDPR compliance is vital to protect necessary information during interactions.

\subsection{Accuracy}

Accuracy requires that personal information collected from system users should be correct. If the information collected is time-sensitive, it is important to ensure that it has not become out-of-date. Maintaining accuracy is crucial not only to ensure the integrity of the system but also for the trust users have in the system. Implementing robust mechanisms to verify data that is used by attesting entities to create micro-credentials after performing verification or creating and submitting transactions on the chain is correct and up-to-date is essential. Regular auditing of information on the blockchain is also required to verify and confirm that data on the chain meets all ethical standards, making it reliable and trustworthy\cite{lapointe2019blockchain}.

\subsection{Right to be Forgotten}

From an ethical point of view, the 'Right to be Forgotten' allows users to request the deletion or removal of their information when it is no longer necessary or relevant. The right to be forgotten gained importance following a lawsuit that led the European Court of Justice to rule that individuals could request search engines like Google to remove their irrelevant personal information from databases and search results \cite{googleCase}. It is usually assumed that the immutability of information on blockchain completely hinders the right to be forgotten. However, this depends on the amount of information and type of information written into it. Selecting a minimal set of required identifiers and storing them on the chain can mitigate the concern about the right to be forgotten \cite{lapointe2019blockchain} to an extent. This is also applicable for decentralized registries maintained as part of an SSI-based system. Right to erasure does not apply to SSI-based verifiable credentials since the controller of the credential is the individual themselves \cite{kondova2020self}.

\subsection{Data Access and Ownership}

In identity systems, access encompasses user interaction, the extent and timing of information access, authority over data access permissions, and mechanisms securing this access \cite{lapointe2019blockchain}. The transparent nature of blockchain does not expose any private information belonging to the user since the proposed system stores only non-sensitive details on the chain. The SSI-based system empowers individuals with control over their own data. Data ownership addresses the questions about who owns the data used in the system for various processes. In the proposed system, users have authority over data generated and issued as credentials to them. Since multiple organizations across countries can use the data stored on the chain to verify data on the attestation chain, a consensus group can be formed to ensure that the data is processed responsibly.

\subsection{Governance}

Governance involves organization-level decisions on who is responsible for different components of a system and decisions taken by the system. In a blockchain-based system, governance is shaped by its centralized or decentralized authority structure and the degree of automation in decision-making. In our proposed system, which runs alongside the current semi-automated attestation process, the attesting entity manually verifies documents before deciding on their legitimacy. Decisions will be based on the existing government policies and regulations. The inclusion of references to policies that led to their decisions can be recorded in the blockchain transaction to enhance transparency and trust.

\section{Conclusion and Future Work}

We introduce a document verification system utilizing blockchain and self-sovereign identity technology to enhance transparency, traceability, and tracking of the verification process from inception to completion. SSI technology ensures privacy for involved parties and secures information used in decision-making. Smart contracts, combined with blockchain's tamper-evident transparency, help processes such as enforcing pre-defined verification rules and dispute resolution and assist attesting entities in generating transactions.

\noindent
Future work involves conducting security risk assessments before the development of the system. A thorough security risk assessment should identify and evaluate potential risks that can compromise the system's security and is crucial to identify parties responsible and liable for the system's actions. This helps develop mitigation strategies to prevent the identified attacks on a system involving blockchain and SSI components. If a technical solution involves any third-party services such as cloud servers, it is important to ensure that all accountable parties reach a consensus and sign an agreement regarding the processing of user data \cite{GDPRchecklist}. Security risk assessment should also be performed before deployment to analyze the impact of security measures on the performance of the system in real-world scenarios. 

\noindent
A comprehensive scalability assessment will be conducted as part of future work, verifying the capacity of a decentralized registry for storage and handling of DID information of all the roles in the system. The system's scalability will also be assessed for its ability to store non-sensitive attestation details on blockchain. This work should also include user testing: In a controlled environment, users can engage with the system functionalities, such as requests for document verification, attestation, and credential management. Feedback will be gathered systematically to evaluate user experience regarding system friendliness and overall satisfaction. This feedback is crucial to refine the system, identify and rectify usability issues, and improve the functionality of the system. 

\noindent
To further advance our initiative, we are working on a proof of concept implementation to validate the system's functionality and demonstrate how blockchain and SSI can effectively streamline the document verification process. By developing this proof of concept, we aim to identify any potential challenges and refine our approach to address them effectively. This hands-on testing will also provide invaluable insights that will help in scaling the system for broader use.

\bibliographystyle{IEEEtran}
\bibliography{mybibfile}

\section*{Authors}
\noindent {\bf Swapna Krishnakumar Radha} received her Master's degree in Computer Science from Indiana University South Bend and a Bachelor's degree in Information Technology. Currently, she is pursuing her PhD in Computer Science at the University of Notre Dame. Her research focus involves using decentralized technologies like blockchain and self-sovereign identity to enhance security, privacy, and trust in various digital systems.\\

\noindent {\bf  Andrey Kuehlkamp} is a research assistant professor in the Center for Research Computing at the University of Notre Dame. His background includes biometrics, computer vision, machine learning, and distributed ledger technology. His research interests reside in the intersection between biometrics, machine learning and distributed systems. He has published a number of research papers in respected peer-reviewed journals and conferences in the field.\\

\noindent {\bf Jarek Nabrzyski} Is a director of the Center for Research Computing at the University of Notre Dame, and a concurrent professor in the Department of Computer Science and Engineering. His lab's research focuses on decentralized systems. \\

\end{document}